# Chapter Number

# Graphene Field Effect Transistors: Diffusion-Drift Theory

G.I. Zebrev
*Department of Micro and Nanoelectronics,*
*National Research NuclearUniversity,*
*"MEPHI", Moscow,*
*Russia*

## 1. Introduction

Recently discovered stable monoatomic carbon sheet (graphene) which is comprised of field-effect structures has remarkable physical properties promising nanoelectronic applications (Novoselov, 2004). Practical semiconductor device simulation is essentially based on diffusion-drift approximation (Sze & Ng, 2007). This approximation remains valid for graphene field-effect transistors (GFET) due to unavoidable presence of scattering centers in the gate or the substrate insulators and intrinsic phonon scattering (Ancona, 2010). Traditional approaches to field-effect transistors modeling suffer from neglect of the key and indispensible point of transport description – solution of the continuity equation for diffusion-drift current in the channels. This inevitably leads to multiple difficulties connected with the diffusion current component and, consequently, with continuous description of the I-V characteristics on borders of operation modes (linear and saturation, subthreshold and above threshold regions). Many subtle and/or fundamental details (difference of behaviour of electrostatic and chemical potentials, specific form of the Einstein relation in charge-confined channels, compressibility of 2D electron system, etc.) are also often omitted in device simulations. Graphene introduces new peculiar physical details (specific electrostatics, crucial role of quantum capacitance etc.) demanding new insights for correct modeling and simulation (Zebrev, 2007). The goal of this chapter is to develop a consequent diffusion-drift description for the carrier transport in the graphene FETs based on explicit solution of current continuity equation in the channels (Zebrev, 1990) which contains specific and new aspects of the problem. Role of unavoidable charged defects near or at the interface between graphene and insulated layers will be also discussed.
Distinguishing features of approach to GFET operation modeling will be:

- diffusion-drift approach;
- explicit solution of current continuity equation in GFET channels;
- key role of quantum capacitance in the diffusion to drift current ratio and transport in GFETs;
- role of rechargeable near-interfacial defects and its influence on small-signal characteristics of GFETs.

## 2. General background

### 2.1 Carrier statistics in ideal graphene for nonzero temperature

The density of states is the number of discrete eigenenergy levels within a unit energy width per unit area (states/eV cm[2]). Taking into account valley and spin as well as angular degeneracy we have for two-dimensional density of states $g_{2D}(\varepsilon)$ in graphene

$$g_{2D}(\varepsilon)d\varepsilon = 4\frac{dp_x dp_y}{(2\pi\hbar)^2} = \frac{4}{(2\pi\hbar)^2}\frac{\varepsilon}{v_0^2}2\pi\, d\varepsilon \,, \tag{1}$$

and specifically for gapless graphene dispersion law $\varepsilon = v_0\sqrt{p_x^2 + p_y^2}$

$$g_{2D}(\varepsilon) = \frac{2\varepsilon}{\pi\hbar^2 v_0^2}\operatorname{sgn}\varepsilon = \frac{2|\varepsilon|}{\pi\hbar^2 v_0^2}\,, \tag{2}$$

where ≜ is the Plank constant, $v_0$ (≅ 10[8] cm/s) is the characteristic (Fermi) velocity in graphene. Using the equilibrium Fermi-Dirac function $f_{FD}(\varepsilon - \mu)$ the electron density per unit area $n_e$ at a given chemical potential $\mu$ for nonzero temperature $T$ reads

$$\begin{aligned} n_e(\mu) &= \int_0^{+\infty} d\varepsilon\, g_{2D}(\varepsilon) f_{FD}(\varepsilon - \mu) = \\ &= \frac{2(k_B T)^2}{\pi\hbar^2 v_0^2}\int_0^{+\infty} du \frac{u}{1 + \exp\left(u - \frac{\mu}{k_B T}\right)} = -\frac{2}{\pi}\left(\frac{k_B T}{\hbar v_0}\right)^2 Li_2\left(-e^{\frac{\mu}{k_B T}}\right), \end{aligned} \tag{3}$$

where T is absolute temperature, $k_B$ is the Boltzmann constant, $Li_n(x)$ is the poly-logarithm function of n-th order (Wolfram, 2003)

$$Li_n(z) = \sum_{k=1}^{\infty} z^k / k^n \tag{4}$$

Using electron-hole symmetry $g(\varepsilon) = g(-\varepsilon)$ we have similar relationship for the hole density $n_h$

$$n_h(\mu) = \int_{-\infty}^{0} d\varepsilon\, g_{2D}(\varepsilon)\left(1 - f_{FD}(\varepsilon - \mu)\right) = -\frac{2}{\pi}\left(\frac{k_B T}{\hbar v_0}\right)^2 Li_2\left(-e^{-\frac{\mu}{k_B T}}\right) = n_e(-\mu)\,. \tag{5}$$

Full charge density per unit area or the charge imbalance reads as

$$n_S \equiv n_e - n_h = \int_0^{+\infty} d\varepsilon\, g_{2D}(\varepsilon)\left(f(\varepsilon - \mu) - f(\varepsilon + \mu)\right) = \frac{2}{\pi}\left(\frac{k_B T}{\hbar v_0}\right)^2 \left(Li_2\left(-e^{-\frac{\mu}{k_B T}}\right) - Li_2\left(-e^{\frac{\mu}{k_B T}}\right)\right). \tag{6}$$

Conductivity of graphene charged sheet is determined by the total carrier density

$$N_S = n_e + n_h = -\frac{2}{\pi}\left(\frac{k_B T}{\hbar v_0}\right)^2 \left( Li_2\left(-e^{-\frac{\mu}{k_B T}}\right) + Li_2\left(-e^{\frac{\mu}{k_B T}}\right)\right). \tag{7}$$

For ideal electrically neutral graphene without any doping (so called the charge neutrality point (NP) with the zero chemical potential $\mu$ = 0) we have intrinsic density with equal densities of electrons and holes

$$N_S(\mu = 0) \equiv 2n_i = -\frac{2}{\pi}\left(\frac{k_B T}{\hbar v_0}\right)^2 2Li_2(-1) = -\frac{2}{\pi}\left(\frac{k_B T}{\hbar v_0}\right)^2 2\sum_{k=1}^{\infty}\frac{(-1)^k}{2^k} = \frac{\pi}{3}\left(\frac{k_B T}{\hbar v_0}\right)^2. \tag{8}$$

Intrinsic carrier density at room temperature T = 300K is estimated to be of order $n_i \cong 8\times10^{10}$ cm$^{-2}$ (slightly larger than in silicon). The Tailor series expansion in the vicinity of the $\mu = 0$

$$-Li_2\left(-e^{\frac{\mu}{k_B T}}\right) \cong \frac{\pi^2}{12} + \ln 2 \frac{\mu}{k_B T} + \frac{\mu^2}{4(k_B T)^2} \tag{9}$$

yields a good approximation for only $\mu < 5k_B T$. It is convenient to use a following asymptotics

$$-Li_2\left(-e^{z}\right) \cong \frac{\pi^2}{12} + \frac{z^2}{2},\ z >> 1; \tag{10}$$

$$-Li_2\left(-e^{\frac{\mu}{k_B T}}\right) \cong \frac{\pi^2}{12} + \frac{\mu^2}{2(k_B T)^2}; \tag{11}$$

$$n_e(\mu) \cong \frac{2}{\pi}\left(\frac{k_B T}{\hbar v_0}\right)^2 \left(\frac{\pi^2}{12} + \frac{\mu^2}{2(k_B T)^2}\right) = \frac{\pi}{6}\left(\frac{k_B T}{\hbar v_0}\right)^2 + \frac{\mu^2}{\pi \hbar^2 v_0^2} = n_i + \frac{\mu^2}{\pi \hbar^2 v_0^2}. \tag{12}$$

This approximation yields both exact expression for electron charge concentration at the charge neutrality point and the correct asymptotics for $\mu >> k_B T$ aw well as good coincidence in the intermediate region $\mu \sim k_B T$. In spite of this fact this approximation is inappropriate for capacitance calculation at zero chemical potential point due to lack of linear terms in $\mu$. In reality the region near the $\mu \sim 0$ should not be considered to be ideal because of inevitable disorder presence (Martin & Akerman, 2008).
The channel electron density per unit area for degenerate system ($\mu >> k_B T$) reads

$$n_S \cong \int_0^{\mu} d\varepsilon\, g_{2D}(\varepsilon) \simeq \frac{\mu^2}{\pi \hbar^2 v_0^2} \tag{13}$$

### 2.2 Quantum capacitance in graphene

Performing explicit differentiation of Eqs.(3,5) one reads

$$\frac{dn_e}{d\mu} = \frac{2}{\pi}\frac{k_BT}{\hbar^2 v_0^2}\ln\left(1+\exp\left(\frac{\mu}{k_BT}\right)\right), \quad \frac{dn_h}{d\mu} = -\frac{2}{\pi}\frac{k_BT}{\hbar^2 v_0^2}\ln\left(1+\exp\left(-\frac{\mu}{k_BT}\right)\right). \tag{14}$$

Exact expression for quantum capacitance (Luryi, 1988) of the graphene charge sheet may be defined as

$$C_Q \equiv \int_{-\infty}^{+\infty} g(\varepsilon)\left(-\frac{\partial f_0}{\partial \varepsilon}\right)d\varepsilon = \frac{e\,d(n_e - n_h)}{d\mu} = \frac{2}{\pi}\left(\frac{e^2}{\hbar v_0}\right)\frac{k_BT}{\hbar v_0}\ln\left(2+2\cosh\left(\frac{\mu}{k_BT}\right)\right). \tag{15}$$

Quantum capacitance for unbiased case ( $\mu$ = 0) becomes formally exact ideal form

$$C_{Q\min} = \frac{2\ln 4}{\pi}\left(\frac{e^2}{\hbar v_0}\right)\frac{k_BT}{\hbar v_0}. \tag{16}$$

For a relatively high doping case ($|\mu| >> k_BT$) we have *approximate* relation for quantum capacitance

$$C_Q \cong e^2\frac{dn_S}{d\mu} = \frac{2}{\pi}\left(\frac{e^2}{\hbar v_0}\right)\frac{|\mu|}{\hbar v_0} \tag{17}$$

For total density of free carriers we have relationship, which is valid for any $\mu$

$$\frac{d(n_e + n_h)}{d\mu} = \frac{2}{\pi}\frac{|\mu|}{\hbar^2 v_0^2}. \tag{18}$$

In contrast to Eq. 17 the latter Eq.18 can be considered as an *exact* for ideal graphene for any chemical potential result connected to an exact form of the Einstein relation.

### 2.3 Einstein relation in graphene

Similar to the silicon MOSFETs, the transport properties of graphene are determined by scattering from the charged defects in the gate insulating oxide and from elastic (at least in low-field region) phonons (Das Sarma et al., 2010). The diffusion constant in 2D graphene sheet can be determined through the Fermi velocity $v_0$ and transport relaxation time $\tau_{tr}$ or mean free path $\ell = v_0\tau_{tr}$

$$D = \frac{1}{2}v_0^2\tau_{tr} = \frac{1}{2}v_0\,\ell. \tag{19}$$

Electron and hole mobility $\mu_{e/h}$ can be inferred from the Einstein relation in a following manner ($e$ = $|e|$)

$$\mu_{e/h} = \frac{eD_{e/h}}{n_{e/h}}\frac{dn_{e/h}}{d\mu} \equiv \frac{eD_{e/h}}{\varepsilon_D}, \tag{20}$$

where a diffusion energy introduced (Ando et al. 1982)

$$\varepsilon_D \equiv n_{e/h}\Big/\left(dn_{e/h}/d\mu\right). \tag{21}$$

It is easy to show from Eq. 13 that rather far from the graphene charge neutrality point we have $\varepsilon_D = \varepsilon_F/2$. Bipolar conductivity is expressed formally with Eq.(20) through the sum of electron and hole components

$$\sigma_0 = e\mu_e n_e + e\mu_p n_p = e^2\left(D_e \frac{dn_e}{d\mu} + D_h \frac{dn_h}{d\mu}\right). \tag{22}$$

Using the exact Eq. 18 and the assumption of electron-hole symmetry ($D_e = D_h = D_0$), the total bipolar conductivity reads

$$\sigma_0 = e^2 D_0 \frac{d(n_e + n_p)}{d\mu} = \frac{2e^2}{h}\frac{\varepsilon_F \tau_{tr}}{\hbar} = \frac{2e^2}{h} k_F \ell\,, \tag{23}$$

where the Fermi wavevector is defined through the dispersion law in gapless graphene $\hbar v_0 k_F = \mu \cong \varepsilon_F$. The Einstein relation can be rewritten in an equivalent form via conductivity and quantum capacitance

$$D_0 C_Q = e\mu_0 N_S = \sigma_0 \tag{24}$$

The Einstein relation allows to easily obtain a relation for mobility of graphene carriers in highly doped ($|\mu| >> k_B T$) graphene

$$\mu_0 = \frac{e v_0 \tau_{tr}}{p_F} = \frac{e\ell}{p_F}. \tag{25}$$

Notice that in fact $\ell \propto p_F$ and $\mu_0$ is independent on Fermi energy in graphene.

## 3. GFET electrostatics

### 3.1 Near-interfacial rechargeable oxide traps

It is widely known (particularly, from silicon-based CMOS practice) that the charged oxide defects inevitably occur nearby the interface between the insulated layers and the device channel. Near-interfacial traps (defects) are located exactly at the interface or in the oxide typically within 1-3 nm from the interface. These defects can have generally different charge states and capable to be recharged by exchanging carriers (electrons and holes) with device channel. Due to tunneling exchange possibility the near-interfacial traps sense the Fermi level position in graphene. These rechargeable traps tend to empty if their level $\varepsilon_t$ are above the Fermi level and capture electrons if their level are lower the Fermi level.

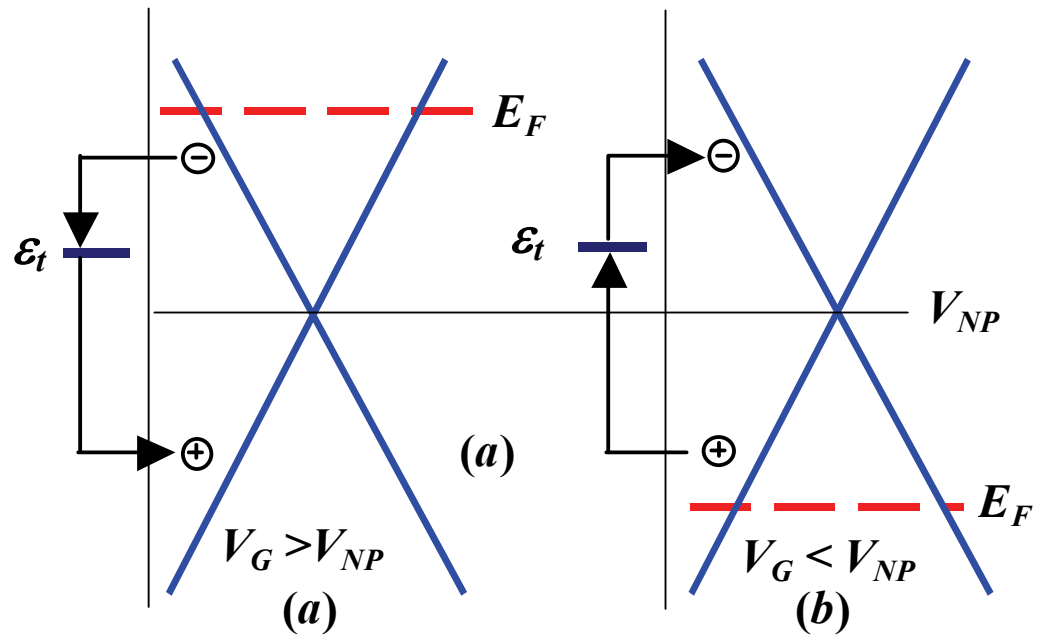

Fig. 1. Illustration of carrier exchange between graphene and oxide defects (a) filling; (b) emptying

There are two types of traps – donors and acceptors. Acceptor-like traps are negatively charged in a filled state and neutral while empty ( - /0). Donor-like traps are positively charged in empty state and neutral in filled condition (0/+). In any case, the Fermi level goes down with an increase $V_G$ and the traps begin filled up, i.e. traps become more negatively charged (see Fig. 1). Each gate voltage corresponds to the respective position of the Fermi level at the interface with own "equilibrium" filling and with the respective density of equilibrium trapped charge $Q_t(\mu) = eN_t(\mu)$ which is assumed to be positive for definiteness. For traps with small recharging time the equilibrium with the substrate would establish fast. These traps rapidly exchanged with the substrate are often referred as to the interface traps ($N_{it}$) (Emelianov et al. 1996); (Fleetwood et al., 2008). Defects which do not have time to exchange charge with the substrate during the measurement time are referred to as oxide-trapped traps ($N_{ot}$). Difference between the interface and oxide traps is relative and depends, particularly, on the gate voltage sweep rate and the measurement's temperature. Interface trap capacitance per unit area $C_{it}$ may be defined in a following way

$$C_{it} \equiv \frac{d}{d\mu}\left(-eN_t(\mu)\right) > 0 \; . \tag{26}$$

Note that the Fermi level dependent $eN_t(\mu)$ contains the charge on all traps, but for a finite voltage sweep time $t_s$ only the "interface traps" with low recharging time constants $\tau_r < t_s$ contribute to the recharging process. Interface trap capacitance (F/cm$^2$) with accuracy up to the dimensional factor represents the energy density of the defect levels $D_{it}$ ( cm$^{-2}$eV$^{-1}$). It is easy to see that these values are related as

$$C_{it} = e^2 D_{it}(\mu) \; . \tag{27}$$

It is useful to note that 1 fF/μm$^2$ ≅ 6.25 × 10$^{11}$ cm$^{-2}$ eV$^{-1}$. The typical interface trap capacitance in modern silicon MOSFETs lies within the range $D_{it}$ ~10$^{11}$ -10$^{12}$ cm$^{-2}$ eV$^{-1}$ and is rather sensitive (especially for thick (> 10 nm) insulated layers) to ionizing radiation impact (Fleetwood et al., 2008).

### 3.2 Electrostatics of graphene gated structures

Let us consider the simplest form of the gate-insulator-graphene (GIG) structure representing the two-plate capacitor capable to accumulate charges of the opposite signs. Without loss of generality we will reference the chemical potential in graphene from the level of charge neutrality $E_{NP}$. Electron affinity (or work function for Dirac point) of graphene with the reference of the vacuum energy level $E_{vac}$ can be defined as

$$\chi_g = E_{vac} - E_{NP} \; . \tag{28}$$

Note that the graphene work function is of order of $\chi_g$ ~ 4.5 eV (Giovannetti et al., 2008). It is well known that voltage bias between any device's nodes is equivalent to applying of electrochemical potential bias. There are generally at least two contributions to the electrochemical potential

$$\mu = \zeta + U = \zeta - e\varphi \tag{29}$$

where $\zeta$ is proper electric charge independent chemical potential, $U$ and $\varphi$ are the electrostatic energy and potential $U = -e\varphi$. Neglecting voltage drop in the gate made routinely of good 3D conductors due to its extremely large quantum capacitance per unit area we get

$$\mu_{gate} = -e\varphi_{gate} - W_{gate}, \tag{30}$$

$$\mu_{graphene} = -\chi_g + \zeta - e\varphi_{graphene} = E_{NP} + \zeta, \tag{31}$$

where $\varphi_{graphene}$ is electrostatic potential of graphene sheet, $W_{gate}$ is work function of the gate material, and $E_{NP} = -\chi_g - e\varphi_{graphene}$ is the energy position of the charge neutrality (or, Dirac) point. Applying the gate voltage (to say, positive) with reference of grounded graphene plate we increase the chemical potential and electrostatic potential of the graphene sheet so as they exactly compensate each other keeping the electrochemical potential of the graphene sample unchanged (see Fig. 2).

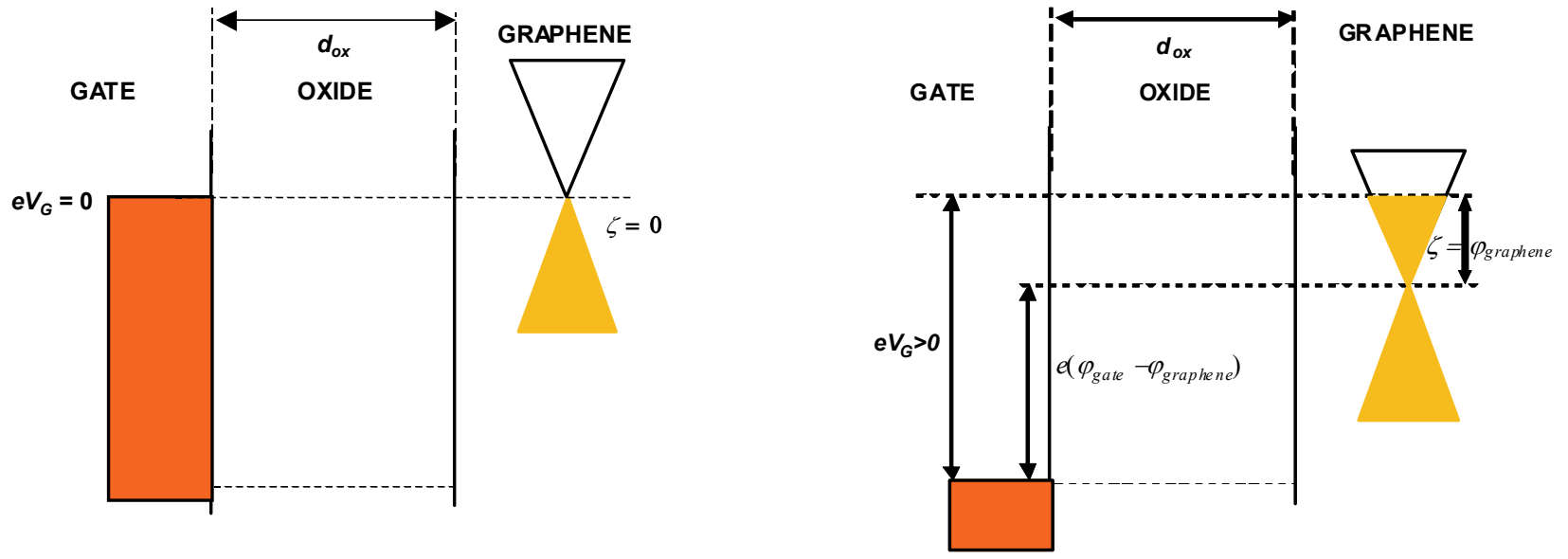

Fig. 2. Band diagram of gate–oxide- graphene structure at $V_G = 0$ (left) and $V_G > 0$ (right). Here, $\varphi_{gg}$ =0, for simplicity.

Particularly, the electrical bias between the metallic (or almost metallic) gate and the graphene sample is equal to a difference between the electrochemical potentials in graphene ( $\mu_{graphene}$ )and the gate ( $\mu_{gate}$ )

$$eV_G = \mu_{graphene} - \mu_{gate} = \varphi_{gg} + \zeta + e\left(\varphi_{gate} - \varphi_{graphene}\right). \tag{32}$$

where $\varphi_{gg} \equiv W_{gate} - \chi_g$ is the work function difference between the gate and graphene. For zero oxide charge (or, for charged oxide defects located nearly the insulator-graphene interface) the electric field $E_{ox}$ is uniform across the gate thickness ($d_{ox}$) and one reads

$$\varphi_{gate} - \varphi_{graphene} = E_{ox}\, d_{ox} = \frac{eN_{gate}}{\varepsilon_{ox}\varepsilon_0} d_{ox} \equiv \frac{eN_{gate}}{C_{ox}}, \tag{33}$$

where $N_{gate}\left(V_G\right)$ is the number of charge carriers on the metallic gate per unit area and the oxide (insulator) capacitance per unit area $C_{ox}$ expressed through the dielectric constants of the insulator ($\varepsilon_{ox}$) is defined as

$$C_{ox} = \frac{\varepsilon_{ox}\varepsilon_0}{d_{ox}}. \tag{34}$$

### 3.3 Characteristic scales of gated graphene

The planar electric charge neutrality condition for the total gated structure can be written down as follows

$$N_G + N_t = n_S \,, \tag{35}$$

where $N_G$ is the number of positive charges per unit area on the gate; $n_S$ is the charge imbalance density per unit area ($n_S$ may be positive or negative and generally non-integer), $N_t$ is the defect density per unit area which is assumed to be positively charged (see Fig.3). Then total voltage drop (Eq.32) across the structure becomes modified as

$$eV_G = \varphi_{gg} + \varphi + \frac{e^2}{C_{ox}}\left(n_S(\zeta) - N_t(\zeta)\right). \tag{36}$$

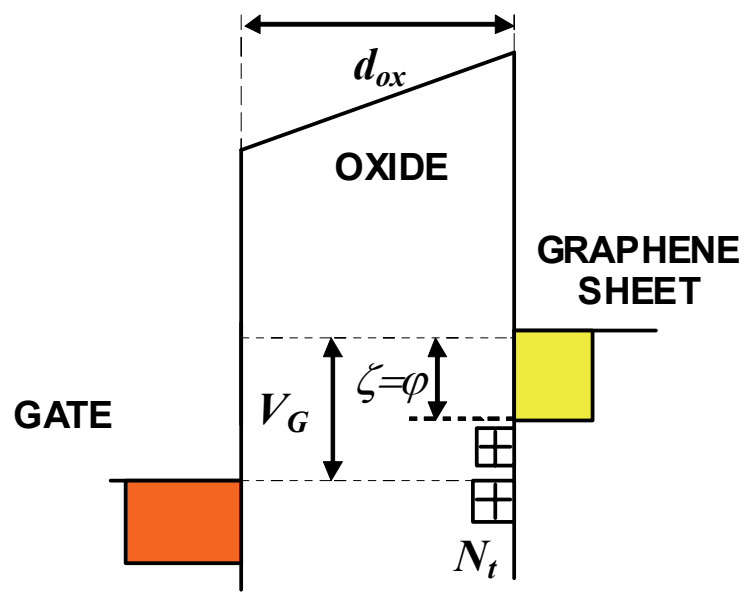

Fig. 3. Band diagram of graphene FET.

The voltage corresponding the electric charge neutrality point gate $V_{NP}$ is defined in a natural way

$$V_{NP} \equiv V_G(\zeta = 0) = \varphi_{gg} - \frac{eN_t(\zeta = 0)}{C_{ox}}. \tag{37}$$

Chemical potential is positive (negative) at $V_G > V_{NP}$ ($V_G < V_{NP}$). Then we have

$$e(V_G - V_{NP}) = \zeta + \frac{e^2 n_S}{C_{ox}} + \frac{e^2\left(N_t(\zeta = 0) - N_t(\zeta)\right)}{C_{ox}}. \tag{38}$$

Taking for brevity without loss of generality $V_{NP}$ =0 and assuming zero interface trap charge at the NP point as well as constant density of trap states we have

$$e^2\left(N_t(\zeta = 0) - N_t(\zeta)\right) \cong C_{it}\,\zeta\,. \tag{39}$$

Taking into account Eq.13 the basic equation of graphene planar electrostatics can be written down a in a form

$$eV_G = \varepsilon_F + \frac{e^2 n_S}{C_{ox}} + \frac{C_{it}}{C_{ox}}\varepsilon_F \equiv m\varepsilon_F + \frac{\varepsilon_F^2}{2\varepsilon_a}, \tag{40}$$

where we have introduced for convenience a dimensionless "ideality factor"

$$m \equiv \ 1+\frac{C_{it}}{C_{ox}}, \tag{41}$$

and notation $\varepsilon_F$ used instead of $\zeta$ . The specificity of the graphene-insulator-gate structure electrostatics is reflected in Eq.40 in appearance of the characteristic energy scale

$$\varepsilon_a = \frac{\pi \hbar^2 v_0^2 C_{ox}}{2e^2} = \frac{\varepsilon_{ox}}{8\alpha_G}\frac{\hbar v_0}{d_{ox}}, \tag{42}$$

where the graphene "fine structure constant" is defined as ( in SI units)

$$\alpha_G = \frac{e^2}{4\pi\varepsilon_0 \hbar v_0}. \tag{43}$$

Fig.4 shows dependencies of characteristic electrostatic energy of gated graphene $\varepsilon_a$ vs gate oxide thickness for typical dielectric constants 4 ($SiO_2$) and 16 ($HfO_2$).

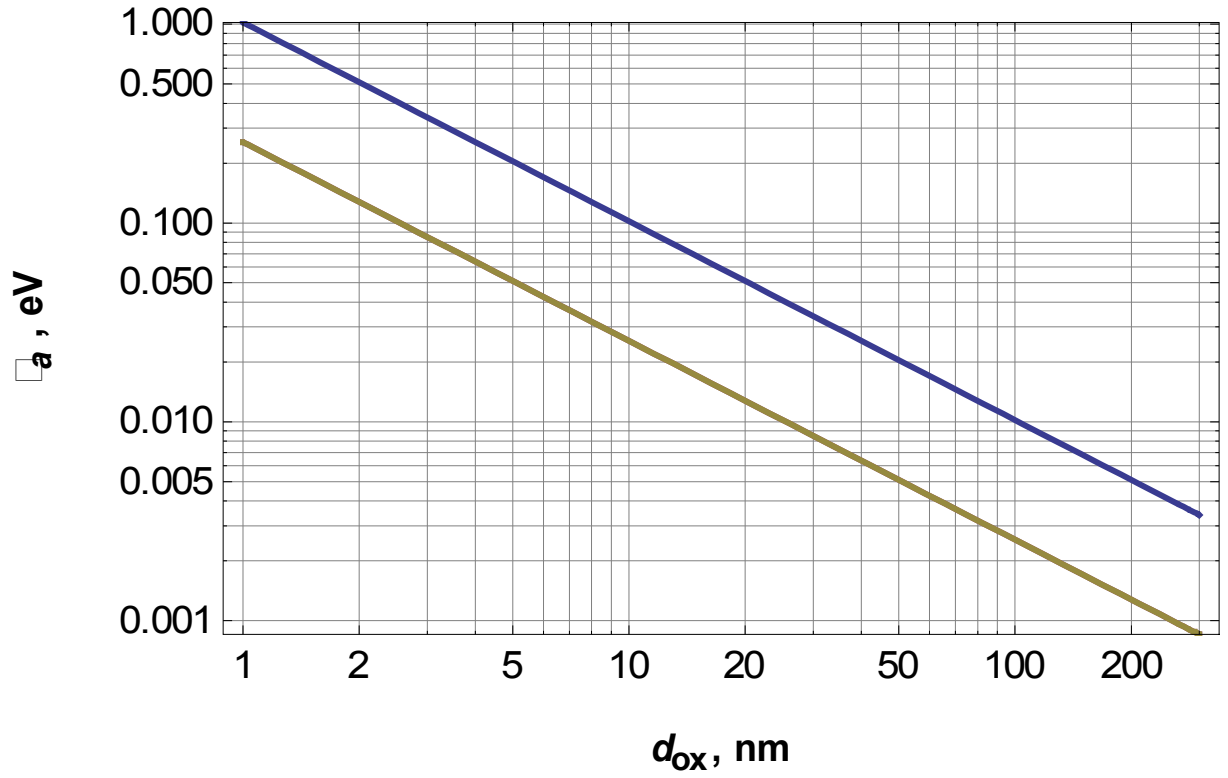

Fig. 4. The dependencies of the $\varepsilon_a$ as functions of the insulator thickness $d_{ox}$ for different dielectric permittivity equal to 4 (lower curve) and 16 (upper curve).

Energy scale $\varepsilon_a$ bring in a natural spatial scale specific to the graphene gated structures

$$a_Q \equiv \frac{\hbar v_0}{\varepsilon_a} = \frac{2e^2}{\pi \hbar v_0 C_{ox}} = \frac{8\alpha_G}{\varepsilon_{ox}} d_{ox}, \tag{44}$$

and corresponding characteristic density

$$n_Q \equiv \frac{1}{\pi a_Q^2} = \frac{\varepsilon_a^2}{\pi \hbar^2 v_0^2} = n_S\left(\varepsilon_F = \varepsilon_a\right). \tag{45}$$

Due to the fact that graphene "fine structure constant" $\alpha_G \cong 2.0-2.2$ the characteristic length $a_Q$ is occasionally of order of the oxide thickness for the insulators with $\varepsilon_{ox}$~16 (i.e. for $HfO_2$). Interestingly that the energy scale $\varepsilon_a$ can be as well represented as functions of the Fermi energy and wavevector $k_F$, quantum capacitance and charge density

$$\frac{\varepsilon_a}{\varepsilon_F} = \frac{C_{ox}}{C_Q} = \frac{1}{k_F a_Q} = \frac{1}{\sqrt{\pi n_S a_Q^2}} \,. \tag{46}$$

### 3.4 Self-consistent solution of basic electrostatic equation

Solving algebraic Eq. (40) one obtains an explicit dependence (to be specific for $V_G > 0$) of the electron Fermi energy as function of the gate voltage

$$\varepsilon_F = \left(m^2\varepsilon_a^2 + 2\varepsilon_a eV_G\right)^{1/2} - m\varepsilon_a \tag{47}$$

This allows to immediately write the explicit relation for graphene charge density dependence on gate voltage

$$\frac{e^2 n_S}{C_{ox}} = eV_G - m\varepsilon_F = eV_G + m^2\,\varepsilon_a - m\left(m^2\varepsilon_a^2 + 2\varepsilon_a eV_G\right)^{1/2} \tag{48}$$

Restoring omitted terms the latter equation can be rewritten as (Zebrev, 2007); (Zebrev, 2007); (Fang et al. 2007)

$$en_S\left(V_G\right) = C_{ox}\left(\left|V_G - V_{NP}\right| + V_0\left(1 - \left(1 + 2\frac{\left|V_G - V_{NP}\right|}{V_0}\right)^{1/2}\right)\right) \tag{49}$$

where the characteristic voltage $V_0 \equiv m^2\varepsilon_a / e$ is defined where interface trap capacitance is taken into account. Figs. 5-6 exhibit numerically the interrelation of $V_0$ with $C_{it}$ and $d_{ox}$ .

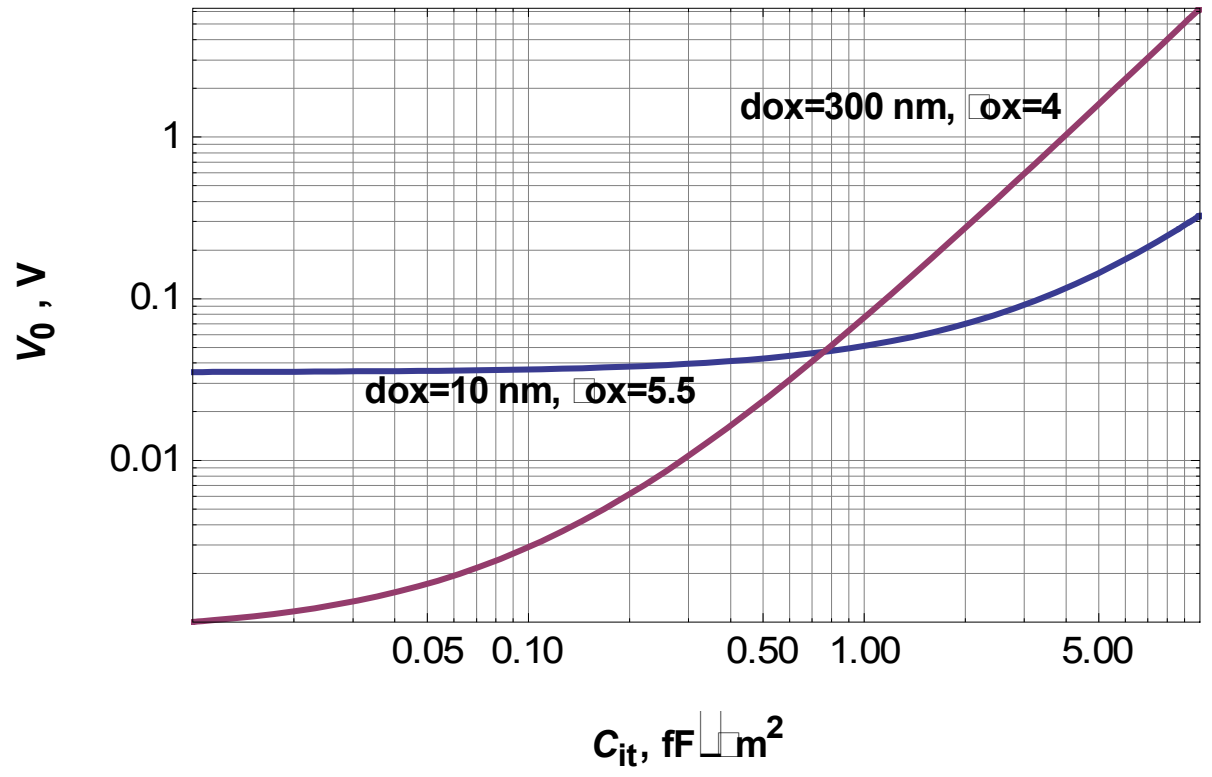

Fig. 5. Simulated dependencies of the characteristic voltage $V_0$ as functions of the interface trap capacitance $C_{it}$ for different oxide parameters.

View of charge density dependence versus gate voltage is determined by relations of characteristic values (see Fig.5,6). At relatively high gate voltage $\left|V_G - V_{NP}\right| >> V_0$ (or, the same, for "thick" oxide) we have close to linear dependence

$$en_S \cong C_{ox}\left(\left|V_G - V_{NP}\right| - \left(2V_0\left|V_G - V_{NP}\right|\right)^{1/2}\right). \tag{50}$$

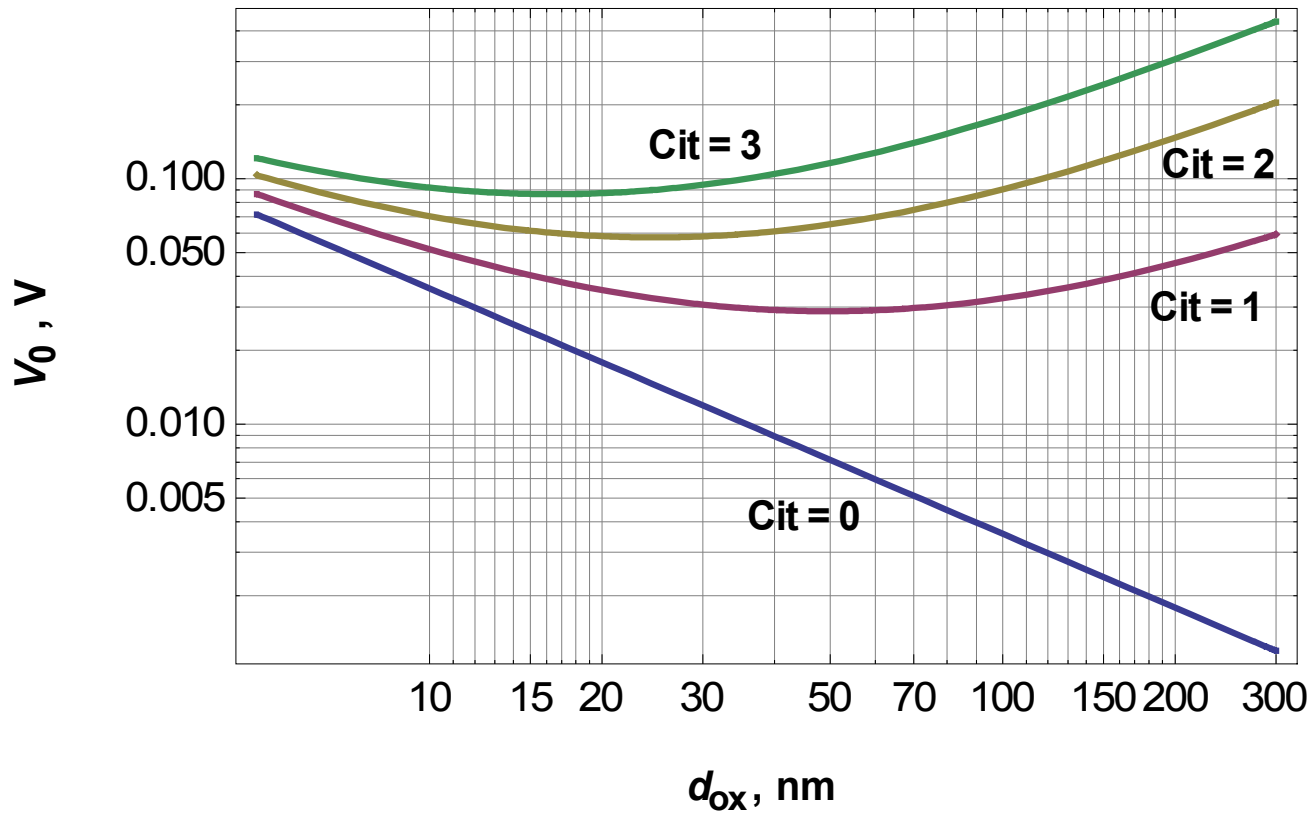

Fig. 6. Simulated dependencies of the characteristic voltage $V_0$ as functions of oxide thickness for different interface trap capacitance (in fF/μm²).

Most part of external gate voltage drops in this case on the oxide thickness. Such is the case of "standard" oxide thickness $d_{ox}$ = 300 nm. Actually for not too small gate bias the charge density dependence on gate voltage is very close to linear (Novoselov et al., 2004). For future graphene FET the gate oxide thickness is assumed to be of order of few or ten of nanometers. For such case of much thinner oxides or under relatively small gate biases $C_{ox}\left|V_G - V_{NP}\right| < en_Q$ we have quadratic law for density dependence (see Fig. 2b)

$$en_S \cong C_{ox}\left(V_G - V_{NP}\right)\left(\frac{V_G - V_{NP}}{V_0}\right), \; V_G - V_{NP} < V_0 . \tag{51}$$

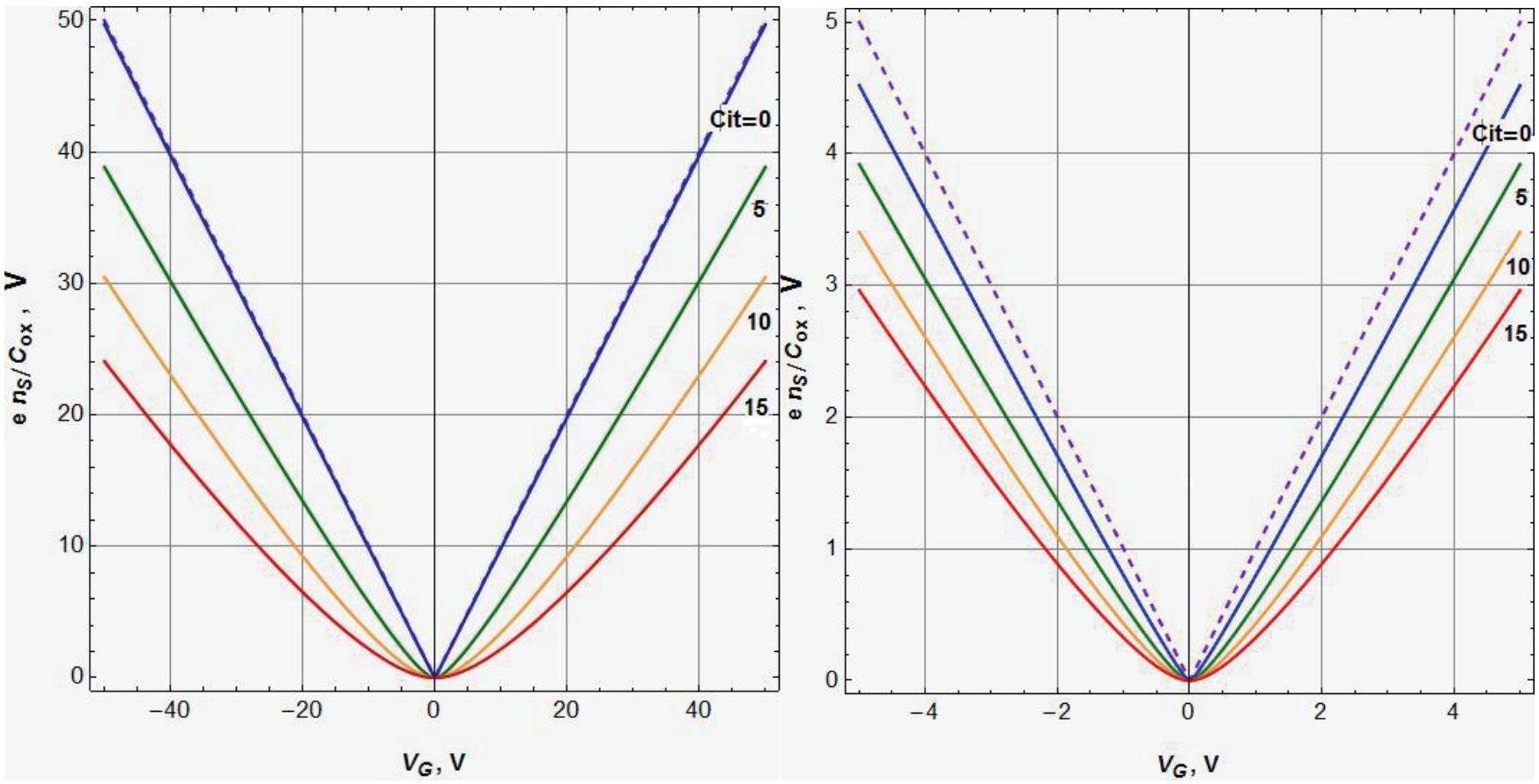

(a) (b)

Fig. 7. Simulated charge density dependencies in reduced form $e\,n_S\,/C_{ox}$ as functions of gate voltage for $\varepsilon_{ox}$ = 4 and different interface trap capacitance $C_{it}$ = 0, 5, 10, 15 *f*F/μm²;
(a) $d_{ox}$ = 300 nm; (b) $d_{ox}$ = 10 nm. Dashed curves correspond to $en_S/C_{ox} = V_G$.

Fig. 7 show that $n_S(V_G)$ curves are strongly affected by interface trap recharging even for relatively thin oxides.

### 3.5 Gate and channel capacitance

Capacitance-voltage measurements are very important in providing information about gated field-effect structures. Taking derivative of Eq. 36 with respect to chemical potential, we have

$$\frac{dV_G}{d\mu} = 1 + \frac{C_Q + C_{it}}{C_{ox}}. \quad (52)$$

Low-frequency gate capacitance can be defined as

$$C_G = e\left(\frac{\partial N_G}{\partial V_G}\right) = e\frac{dN_G/d\mu}{dV_G/d\mu} = \frac{C_Q + C_{it}}{1 + \frac{C_Q + C_{it}}{C_{ox}}} = \left(\frac{1}{C_{ox}} + \frac{1}{C_Q + C_{it}}\right)^{-1} \quad (53)$$

This relation corresponds to the equivalent electric circuit which is shown in Fig.8.

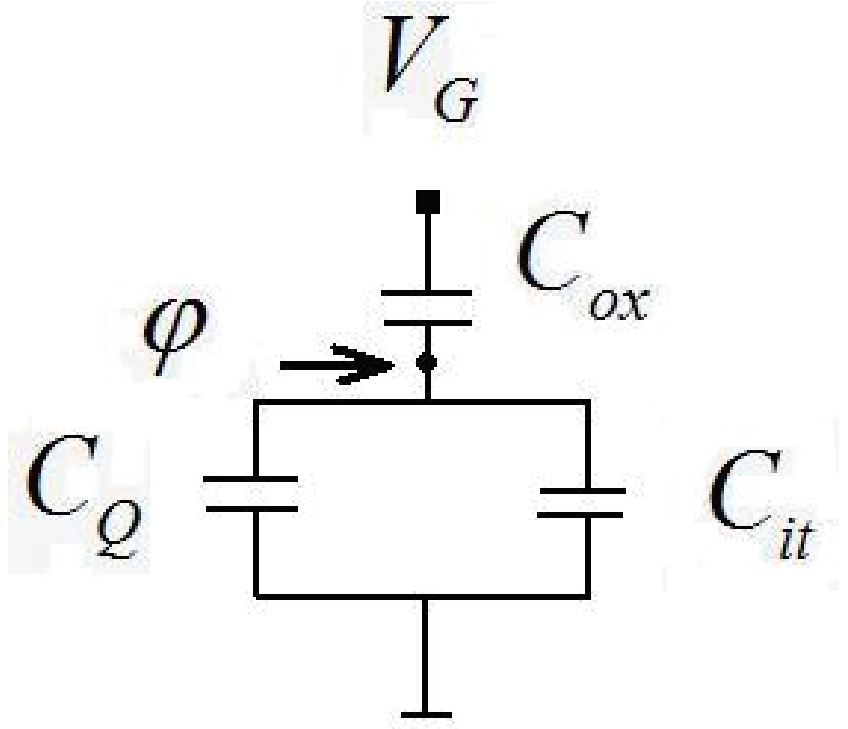

Fig. 8. Equivalent circuit of gated graphene.

One might introduce another relation corresponding to the intrinsic channel capacitance

$$C_{CH} = e\left(\frac{\partial N_S}{\partial V_G}\right) = e\frac{dN_S/d\mu}{dV_G/d\mu} = \frac{C_Q}{1 + \frac{C_Q + C_{it}}{C_{ox}}} = \frac{C_{ox}}{1 + \frac{C_{ox} + C_{it}}{C_Q}}. \quad (54)$$

where all capacitances are non-zero and assumed to be positive values for any gate voltage. Note that $C_{CH}$ is often referred to as "total gate capacitance $C_{tot}$" in literature wherein the interface trap capacitance is frequently ignored. The gate and the channel capacitances are connected in graphene gated structures through exact relation

$$\frac{C_G}{C_{CH}} = 1 + \frac{C_{it}}{C_Q} \quad (55)$$

and can be considered to be coincided only for ideal devices without interface traps when $C_{it}$ =0. All relationships for the differential capacitances remain valid for any form of interface trap energy spectrum. In an ideal case capacity-voltage characteristics $C_{CH}(V_G)$ should be symmetric with refer to the neutrality point implying approximately flat energy density spectrum of interface traps. For the latter case the channel capacity can be derived by direct differentiation of explicit dependence $n_S(V_G)$ in Eq.49

$$C_{CH} = e\frac{dn_S}{dV_G} = C_{ox}\left[1 - \frac{1}{\left[1 + 2\left|V_G - V_{NP}\right|/V_0\right]^{1/2}}\right] \tag{56}$$

As can be seen in Fig.9 the capacitance-voltage characteristics $C_G(V_G)$ is strongly affected by the interface trap capacitance.

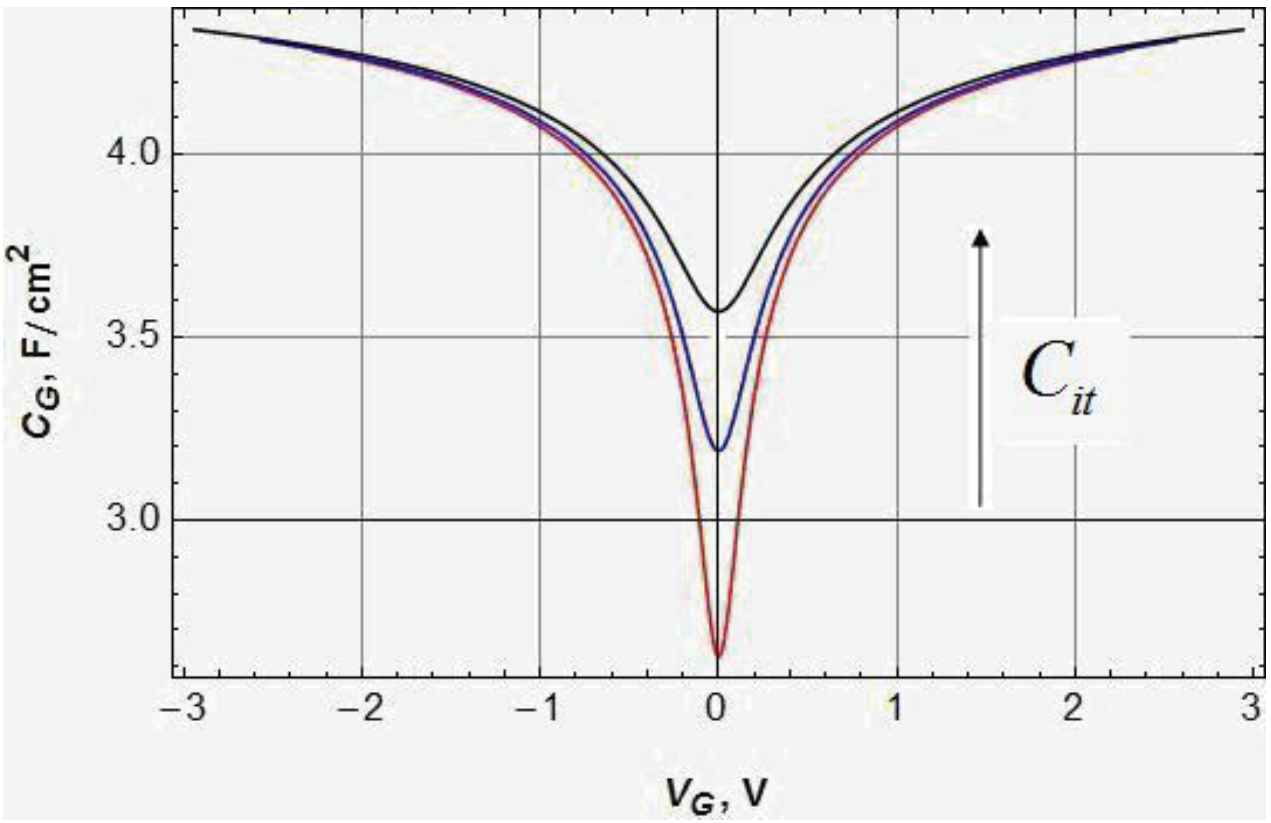

Fig. 9. Simulated dependencies of the gate capacitance $C_G(V_G)$ for different $C_{it}$ = 1, 5, 10 *f*F/µm²; $d_{ox}$ = 10 nm, $\varepsilon_{ox}$ = 5.5 ($Al_2O_3$).

For the case $C_{it}$ = 0 (i.e. m = 1) capacitance-voltage dependencies can be considered as to be universal curves depending on only thickness and permittivity of the gate oxide through the parameter $\varepsilon_a$. In practice one should discriminate the quantum and the interface trap capacitances and this is a difficult task since they are in a parallel connection in equivalent circuit. Comparison of "ideal" capacitance –voltage characteristics with real measured ones represent a standard method of interface trap spectra parameter extraction (Sze & Ng, 2007, Chap. 4,); ( Nicollian & Brews, 1982).

## 4. Diffusion drift current in graphene channels

### 4.1 Diffusion to drift current ratio

It is well-known that the channel electron current per unit width $J_S$ can be expressed as a sum of drift and diffusion components

$$J_S = J_{DR} + J_{DIFF} = e\mu_0 n_S \frac{d\varphi}{dy} + eD_0 \frac{dn_S}{dy}\,, \tag{57}$$

where $\mu_0$ and $D_0$ are the electron mobility and diffusivity, $y$ is a coordinate along the channel. This one can be rewritten in an equivalent form

$$J_D = \sigma_0 E\left(1 - \left(\frac{D_0}{\mu_0}\right)\left(\frac{dn_S}{n_S d\zeta}\right)\left(\frac{d\zeta}{d\varphi}\right)\right), \tag{58}$$

where $E = -d\varphi / dy$ is electric field along the channel, $\sigma_0 = e\mu_0 n_S$ is the graphene sheet conductivity, $\zeta(y)$ and $\varphi(y)$ are the local chemical and electrostatic potential in the graphene channel , respectively. Using the Einstein relation for 2D system of non-interacting carrier as in Eq. 20 the diffusion-drift current reads (Zebrev & Useinov, 1990)

$$J_S \equiv e\mu_0 n_S E\left(1 - \frac{d\zeta}{ed\varphi}\right) = e\mu_0 n_S E\,(1+\kappa). \tag{59}$$

The ratio of the diffusion to the drift current is introduced in Eq.59 as the ratio of gradients of chemical ($\zeta$) and electrostatic ($\varphi$) potentials along the channel, which are the components of electrochemical potential (or local Fermi energy for high doping case)

$$\kappa \equiv -\frac{d\zeta}{ed\varphi} = \frac{J_{DIF}}{J_{DR}} \tag{60}$$

Note that for equilibrium case the electrochemical potential is position independent ($\mu = \zeta - e\varphi$ =const) and $d\zeta / d\varphi$ is identically equals to unity and diffusion-drift current components exactly compensate each other

$$\left(\frac{\partial\zeta}{e\partial\varphi}\right)_\mu = -\frac{(\partial\mu/\partial\varphi)_\zeta}{e(\partial\mu/\partial\zeta)_\varphi} = 1\,. \tag{61}$$

On the contrary for non-equilibrium case both diffusion-drift components have the same direction ($d\zeta / d\varphi < 0$) and the parameter $\kappa > 0$. Unlike to the equilibrium case the electrostatic and chemical potential should considered as independent variables in non-equilibrium systems; e.g., the chemical potential controls particle (electron) density and is generally irrelevant to properly electric charge density and electrostatic potential. Two-dimensional electron density in the channel $n_S(\zeta)$ is a function exactly of the local chemical potential $\zeta$ rather than electrostatic ($\varphi$) or total electrochemical potential ($\mu$). It is very important that the electrochemical potential distribution along the channel does not coincide in general with electrostatic potential distribution.

To properly derive explicit expression for control parameter $\kappa$ we have to use the electric neutrality condition along the channel length in gradual channel approximation which is assumed to be valid even under non-equilibrium condition $V_{DS} > 0$. Differentiating Eq.36 with respect to chemical potential $\zeta$ (note that $V_G = const(y)$) and taking into consideration that $\varphi(y)$ and $\zeta(y)$ in the channel are generally non-equal and independent variables and $n_S$ depends on only chemical potential $\zeta$ one can get

$$\kappa = -\left(\frac{\partial\zeta}{e\partial\varphi}\right)_{V_G} = \frac{(\partial V_G/\partial\varphi)_\zeta}{e(\partial V_G/\partial\zeta)_\varphi} = \frac{C_{ox}}{C_Q + C_{it}} \tag{62}$$

This dimensionless parameter $\kappa$ is assumed to be constant along the channel for a given electric biases and expressed via the ratio of characteristic capacitances. For ideal graphene channel with low interface trap density the $\kappa$-parameter is a function of only $\varepsilon_a$ and the Fermi energy

$$\kappa(C_{it}=0)=\frac{C_{ox}}{C_Q}=\frac{\varepsilon_a}{\varepsilon_F}=\frac{1}{k_F a_Q}=\frac{1}{\sqrt{\pi n_S a_Q^2}}. \qquad (63)$$

For a high-doped regime (large $C_Q$) and/or thick gate oxide (low $C_{ox}$) when $C_Q >> C_{ox}$ we have $\kappa << 1$ by this is meant that the drift current component dominate the diffusion one and vice versa.

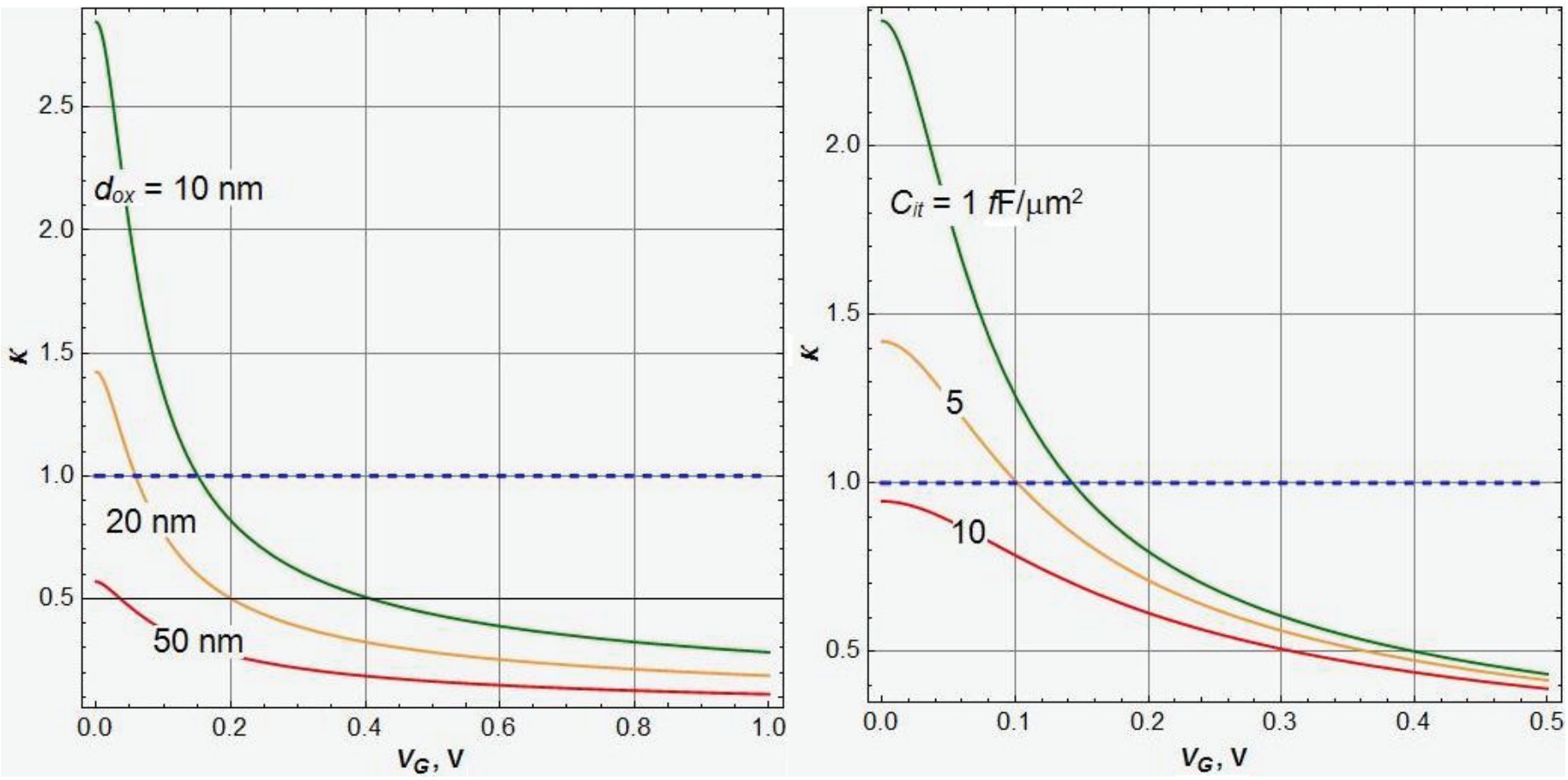

(a) (b)

Fig. 10. Simulated $\kappa$ curves as functions of gate voltage (a) for different oxide thicknesses, $C_{it}$ = 0, $\varepsilon_{ox}$ = 16; (b) for different interface trap capacitances $C_{it}$ = 1, 5, 10 fF/μm², $\varepsilon_{ox}$ = 16, $d_{ox}$ = 10 nm.

Fig. 10 shows simulated dependencies of the parameter $\kappa$ on gate voltage at variety of parameters.

### 4.2 Current continuity equation

The key point of this approach is an explicit analytical solution of continuity equation for channel current density. Total drain current $J_S = J_{DR} + J_{DIFF}$ should be conserved along the channel

$$\frac{dJ_S}{dy}=0 \;\Leftrightarrow\; \frac{d}{dy}(n_S E)=0 \qquad (64)$$

that yields an equation for electric field distribution along the channel (Zebrev & Useinov, 1990)

$$\frac{dE}{dy} = \left(\frac{e\,dn_S}{n_S d\zeta}\right)\left(-\frac{d\zeta}{e d\varphi}\right)\left(\frac{d\varphi}{dy}\right) = \frac{\kappa e^2}{\varepsilon_D}E^2 . \tag{65}$$

where $\kappa$ and $\varepsilon_D$ are assumed to be functions of only the gate voltage rather than the drain-source bias and position along the channel. Direct solution of ordinary differential Eq. 65 yields

$$E(y) = \frac{E(0)}{1 - \dfrac{\kappa e^2 E(0)}{\varepsilon_D} y} , \tag{66}$$

where $E(0)$ is electric field near the source, which should be determined from the condition imposed by a fixed electrochemical potential difference between drain and source $V_D$, playing a role of boundary condition

$$V_D = (1+\kappa)\int_0^L E(y)\,dy , \tag{67}$$

where $L$ is the channel length. Using Eqs. (66) and (67) one obtains an expressions for $E(0)$ and electric field distribution along the channel

$$E(0) = \frac{\varepsilon_D/e}{\kappa L}\left(1 - \exp\left(-\frac{\kappa}{1+\kappa}\frac{eV_D}{\varepsilon_D}\right)\right); \tag{68}$$

$$E(y) = \frac{\dfrac{\varepsilon_D/e}{\kappa L}\left(1 - \exp\left(-\dfrac{\kappa}{1+\kappa}\dfrac{eV_D}{\varepsilon_D}\right)\right)}{1 - \dfrac{y}{L}\left(1 - \exp\left(-\dfrac{\kappa}{1+\kappa}\dfrac{eV_D}{\varepsilon_D}\right)\right)}. \tag{69}$$

### 4.3 Distributions of chemical and electrostatic potential along the channels

Integrating Eq. (69) we have obtained the explicit relationships for distributions of the chemical and electrostatic potentials along the channel length separately and electrochemical potential as a whole

$$\varphi(y) - \varphi(0) = -\frac{\varepsilon_D}{\kappa e} ln\left[1 - \frac{y}{L}\left[1 - \exp\left(-\frac{\kappa}{1+\kappa}\frac{eV_D}{\varepsilon_D}\right)\right]\right], \tag{70}$$

$$\zeta(y) - \zeta(0) = \varepsilon_D\, ln\left[1 - \frac{y}{L}\left[1 - \exp\left(-\frac{\kappa}{1+\kappa}\frac{eV_D}{\varepsilon_D}\right)\right]\right], \tag{71}$$

$$\mu(y) = \mu(0) + \varepsilon_D \frac{1+\kappa}{\kappa} ln\left[1 - \frac{y}{L}\left[1 - \exp\left(-\frac{\kappa}{1+\kappa}\frac{eV_D}{\varepsilon_D}\right)\right]\right], \tag{72}$$

where $\zeta(0)$, $\mu(0)$ and $\varphi(0)$ are the potentials nearby the source controlled by the gate-source bias $V_{GS}$. For any gate voltage $V_{GS}$ (and corresponding $\kappa(V_G)$) the full drop of electrochemical potential μ on the channel length is fixed by the source-drain bias $V_D$

$$e(\varphi(L)-\varphi(0))+\zeta(0)-\zeta(L)=\frac{eV_{DS}}{1+\kappa}+\frac{e\kappa V_{DS}}{1+\kappa}=eV_{DS} \tag{73}$$

Expanding Eqs. 70 at low drain bias and high carrier density case ($\kappa < 1$) we have familiar linear dependence of electrostatic potential on coordinate (as in any good conductor)

$$\varphi(y)-\varphi(0)\cong\frac{y}{L}V_D, \tag{74}$$

and negligible spatial change in chemical potential along the channel length $\Delta\zeta=\kappa\Delta\varphi<<\varphi$. Thus the full drop of chemical potential is negligible under high-doped channel compared to electrostatic potential but it becomes very important in saturation mode.

## 5. Channel current modeling

### 5.1 Current-voltage characteristics

The total drain current at constant temperature can be written as gradient of the electrochemical potential taken in the vicinity of the source

$$\begin{aligned} I_D &= -W\mu_0 n_S(0)\left(\frac{d\mu}{dy}\right)_{y=0} = eW\mu_0 n_S(0)(1+\kappa)E(0)= \\ &= e\frac{W}{L}D_0 n_S(0)\frac{1+\kappa}{\kappa}\left(1-\exp\left(-\frac{\kappa}{1+\kappa}\frac{eV_D}{\varepsilon_D}\right)\right), \end{aligned} \tag{75}$$

where $W$ is the channel width, and the Einstein relation $D_0=\mu_0\varepsilon_D/e$ is employed. Notice that the total two-dimensional charge density $eN_S\cong en_S$ practically equals to charge imbalance density excepting the vicinity of the charge neutrality point where diffusion-drift approximation is failed.

Let us define the characteristic saturation source-drain voltage $V_{DSAT}$ in a following manner

$$V_{DSAT}=2\frac{1+\kappa}{\kappa}\frac{\varepsilon_D}{e}=\frac{1+\kappa}{\kappa}\frac{\varepsilon_F}{e}, \tag{76}$$

where $\varepsilon_F$ is the Fermi energy (the same chemical potential) nearby the source (recall that $\varepsilon_D\cong\varepsilon_F/2$ for $\zeta=\varepsilon_F>>k_BT$). Notice that employing this notation and Eq.71 one might write the chemical potential nearby the drain as

$$\zeta(L)=(1-V_D/V_{DSAT})\varepsilon_F. \tag{77}$$

This implies that the condition $V_D=V_{DSAT}$ corresponds to zero of the chemical potential and current due to electrostatic blocking which is known as pinch-off for silicon MOSFETs (Sze & Ng, 2007). Actually, one might rewrite a general expression for the channel current as

$$I_D = \frac{W}{L}\sigma_0(0)\frac{V_{DSAT}}{2}\left(1-\exp\left(-2\frac{V_D}{V_{DSAT}}\right)\right) \tag{78}$$

where $\sigma_0$ is the low-field conductivity nearby the source. It is evident from Eq.78 that $V_{DSAT}$ corresponds to onset of drain current saturation. This expression describe I-V characteristics of graphene current in a continuous way in all operation modes (see Fig.11)

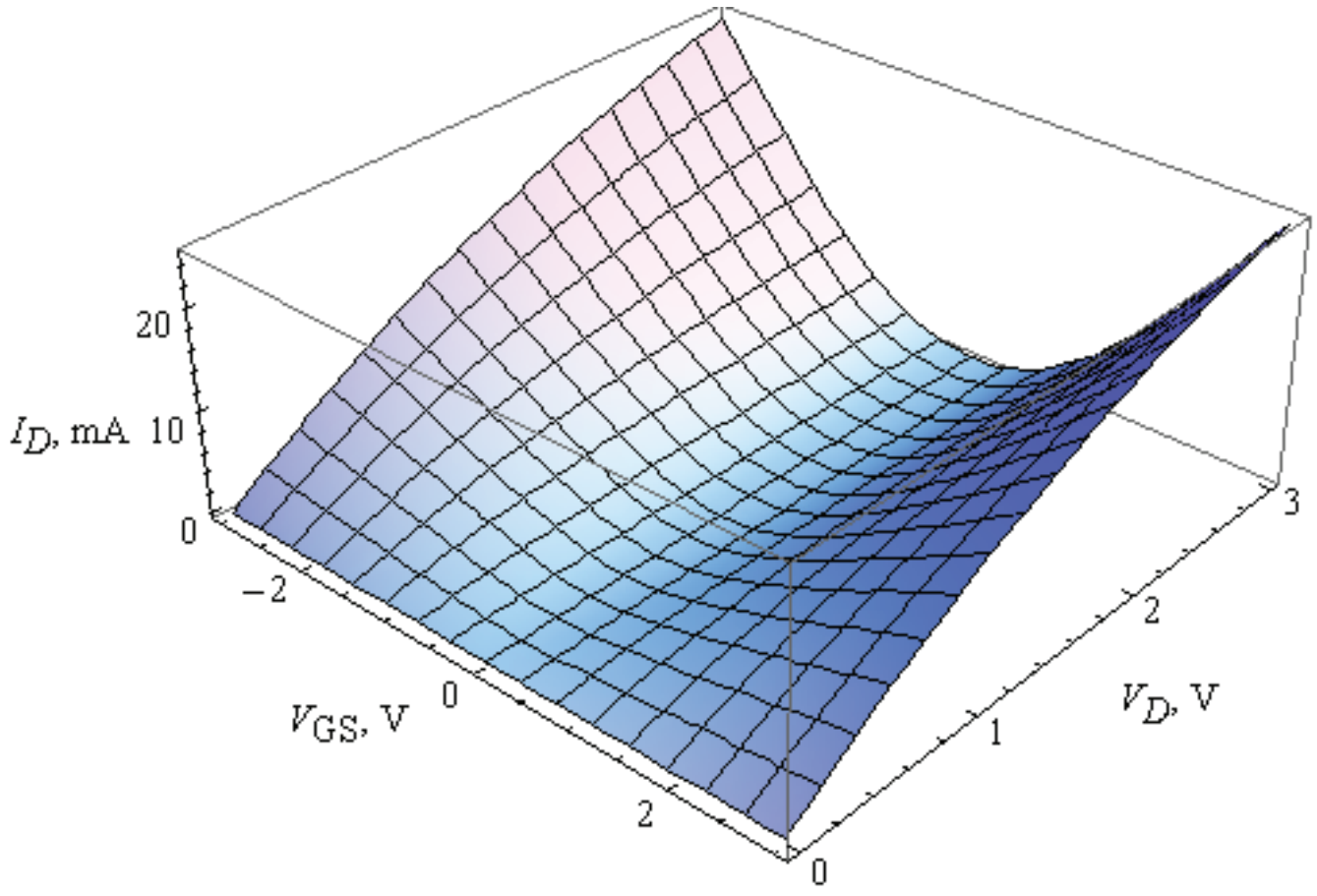

Fig. 11. Current voltage characteristics of graphene FET as function of gate and drain voltage.

### 5.2 Pinch-off (saturation) regime

Taking into account Eqs. 76, 62 and 63 one obtains

$$eV_{DSAT} = \varepsilon_F\frac{1+\kappa}{\kappa} = m\varepsilon_F + \frac{\varepsilon_F^2}{\varepsilon_a}. \tag{79}$$

Recall that $|V_G - V_{NP}| = m\varepsilon_F + \varepsilon_F^2 / 2\varepsilon_a$ one may derive an expression

$$V_{DSAT} = |V_G - V_{NP}| + \frac{\varepsilon_F^2}{2\varepsilon_a} = |V_G - V_{NP}| + \frac{en_S}{C_{ox}}, \tag{80}$$

which is specific for graphene field-effect transistors.

Notice that for thick oxide GFET we have very large $V_{DSAT} \cong 2|V_G - V_{NP}| >> 1$ V and pinch-off saturation is never observed. As can be seen in Fig. 12 the saturation voltage $V_{DSAT}$ depends parametrically on the $\varepsilon_a$ and on interface trap capacitance $C_{it}$. Under condition of high source-drain bias $V_D > V_{DSAT}$ the Eq.78 yields formal relationship for saturation current regime caused by electrostatic pinch-off.

$$I_{DSAT} \cong \frac{W}{L}D_0 n_S(0)\frac{1+\kappa}{\kappa} = \frac{W}{L}\sigma_0(0)\frac{V_{DSAT}}{2} \tag{81}$$

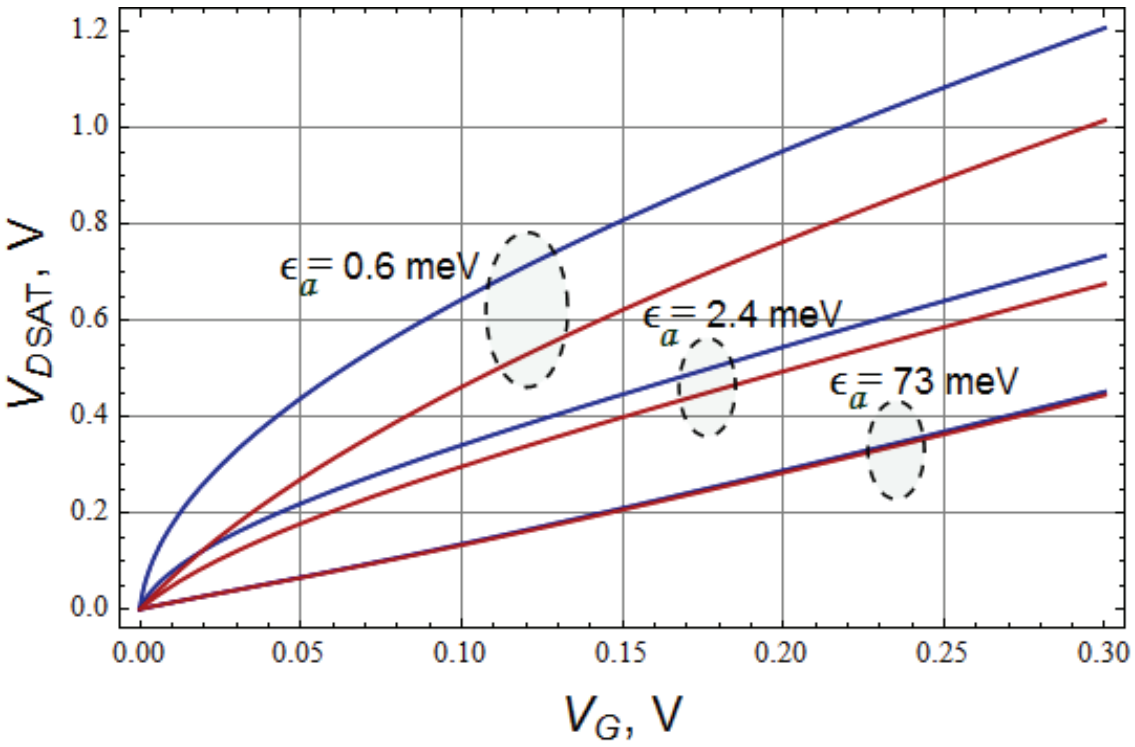

Fig. 12. Simulated $V_{DSAT}$ curves as functions of gate voltage for $\varepsilon_a$ = 0.6 meV ( $d_{ox}$ = 300 nm, $\varepsilon_{ox}$ = 4); $\varepsilon_a$ = 2.4 meV ( $d_{ox}$ = 300 nm, $\varepsilon_{ox}$ = 16); $\varepsilon_a$ = 73 meV ( $d_{ox}$ = 10 nm, $\varepsilon_{ox}$ = 16); $C_{it}$ = 0 *f*F/µm2 (upper curve in the pairs) and $C_{it}$ =1 *f*F/µm2 (lower curve).

### 5.3 Low-field linear regime

Linear (triode) operation mode corresponds to condition

$$V_D << V_{DSAT} = \varepsilon_F \frac{1+\kappa}{\kappa} \, . \tag{82}$$

For high doping regime when $\kappa << 1$ one has predominance of drift component of the channel current as in any metal. In contrast for $\kappa >> 1$ the diffusion current prevails. Equality of the current components occurs in ideal structure $(C_{it} = 0)$ at $\varepsilon_F = \varepsilon_a$ or, equivalently, at the characteristic channel density $n_S = n_Q$, defined in Eq.45.

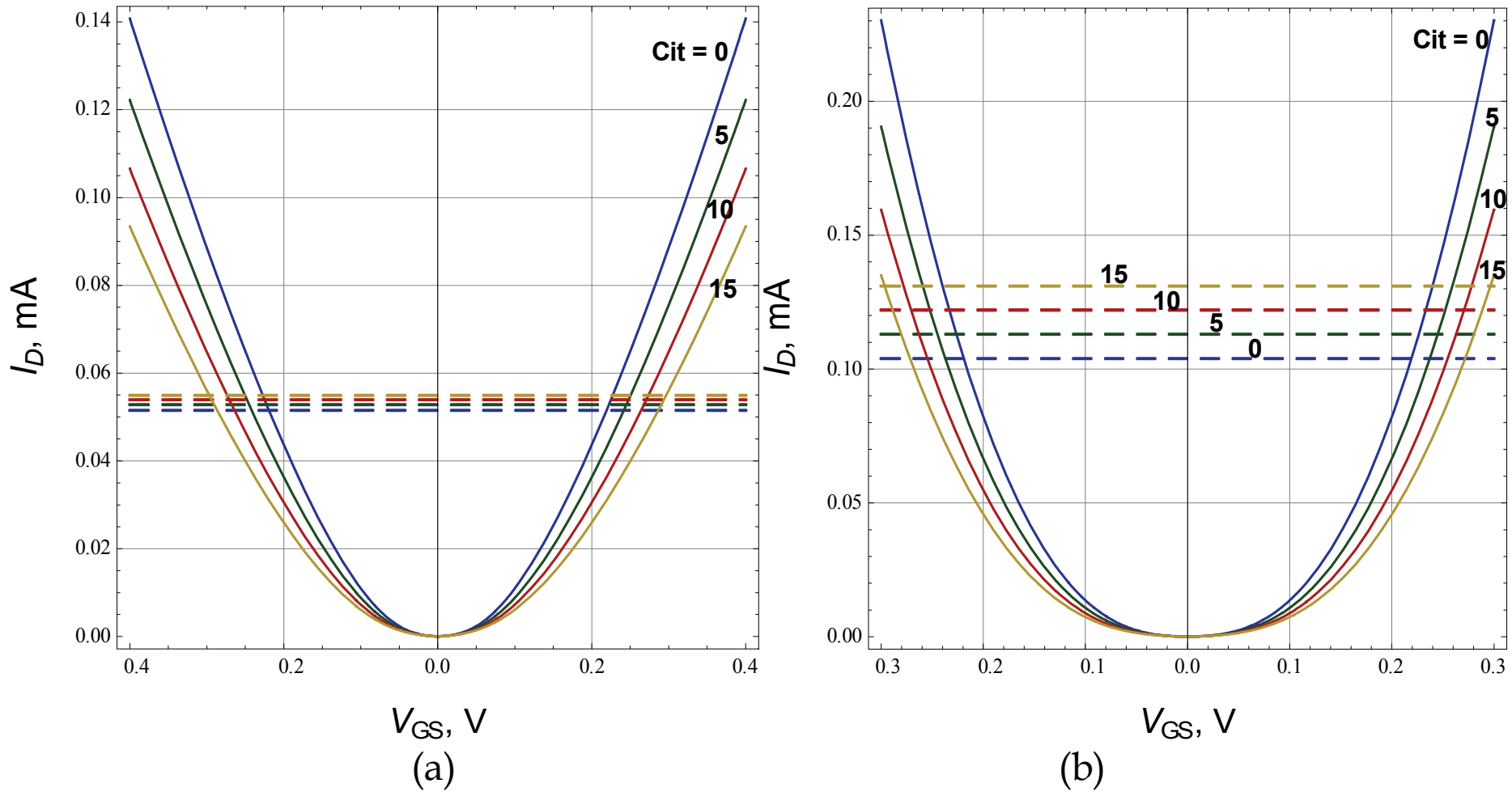

Fig. 13. Simulated drain channel currents as functions of gate voltage for different interface trap capacitances $C_{it}$ = 0, 5, 10, 15 *f*F/µm2; $d_{ox}$ = 5 nm, $\varepsilon_{ox}$ = 16; $W$ = 1 µm; $L$ = 0.25 µm, $\mu_0$ = 800 cm2/(V s); (a) $V_D$ = 0.1 V; (b) $V_D$ = 1 V. Dashed curves correspond to condition $\kappa = 1$.

Fig. 13 shows simulated transfer ($I_D$ vs $V_G$) characteristics of graphene FET for different drain biases and interface trap capacitances. Portions of curves below the dashed curves correspond to predominance of diffusion current with pronounced current saturation, and the above dashed curves correspond mainly to drift current with linear dependence on the drain bias. Notice that the diffusion current region is negligible for dirty structures with thick oxides. For rather small drain bias one can get a usual linear expression expanding Eq.78 in series on $V_D$

$$I_D \cong e\frac{W}{L}\mu_0 n_S V_D . \tag{83}$$

Setting mobility $\mu_0$ gate voltage independent the small-signal transconductance in the linear mode reads

$$g_m \equiv \left(\frac{\partial I_D}{\partial V_G}\right)_{V_D} = \frac{W}{L}\mu_0 C_{CH} V_D , \tag{84}$$

where the channel capacitance $C_{CH}$ is defined in Eq.54. Field-effect mobility $\mu_{FE}$ can be defined from Eq.84 as

$$g_m = \frac{W}{L}\mu_0 V_D C_{CH} \equiv \frac{W}{L}\mu_{FE} V_D C_{ox} . \tag{85}$$

Eq.91 connects field-effect mobility $\mu_{FE}$ depending on charge exchange with extrinsic traps (defects in the gate oxides, chemical dopants etc.) and mobility $\mu_0$ depending on only "microscopic" scattering mechanisms

$$\mu_{FE} = \frac{\mu_0}{1+\dfrac{C_{ox}+C_{it}}{C_Q}} = \frac{\mu_0\ \varepsilon_F}{m\varepsilon_a + \varepsilon_F} . \tag{86}$$

Note that the field-effect mobility, determined often immediately as a slope of the experimental conductivity vs gate voltage curves, is always less than truly microscopic mobility and significantly decreases nearby the charge neutrality point. In fact, $\mu_{FE}$ is close to $\mu_0$ only if $C_Q >> mC_{ox}$ (or, equivalently, $\varepsilon_F >> m\varepsilon_a$), i.e. for a high doping regime.

Transconductance in field-effect transistors commonly degrades affected by electric stress, wear-out or ionizing radiation due to interface trap buildup. The field-effect mobility renormalization after externally induced interface trap capacitance alteration $C_{it} \to C_{it} + \Delta C_{it}$ can be expressed using Eq.86 via initial value $\mu_{FE}(C_{it})$

$$\mu_{FE}(C_{it}+\Delta C_{it}) = \frac{\mu_{FE}(C_{it})}{1+\dfrac{\Delta C_{it}}{C_{ox}+C_Q+C_{it}}} . \tag{87}$$

Logarithmic swing which characterizes the $I_{ON}/I_{OFF}$ ratio and equals numerically to the gate voltage alteration needed for current change by an order can be computed using Eq.83 and Eq.54 as

$$S \equiv \left( \frac{d(\log_{10} I_D)}{dV_G} \right)^{-1} = \ln 10 \left( \frac{dn_S}{n_S dV_G} \right)^{-1} = \ln 10 \, \frac{en_S}{C_{CH}(V_G)} . \quad (88)$$

This formula can be written down in a form more familiar from silicon MOSFET theory

$$S = \ln 10 \left( \frac{en_S}{C_Q} \right) \left( 1 + \frac{C_{it} + C_Q}{C_{ox}} \right) = \ln 10 \left( \frac{en_S}{C_{ox}} \right) \left( 1 + \frac{C_{it} + C_{ox}}{C_Q} \right) = \ln 10 \, en_S \left( \frac{m}{C_Q} + \frac{1}{C_{ox}} \right) . \quad (89)$$

Recall that the diffusion energy $\varepsilon_D = e^2 n_S / C_Q \cong \varepsilon_F / 2$ plays here role of the thermal potential $e\varphi_D = k_B T$ for the subthreshold (non-degenerate) operation mode of the silicon FETs wherein $C_Q$ is negligible. Unlike the silicon FET case the subthreshold swing is a function of gate voltage. Excluding a small region nearby the Dirac point the latter expression yields an assessment of the logarithmic swing $S \geq \ln 10 \, en_S / C_{ox} >>$ 1V/decade for "thick" oxides and "clean" interface ($C_Q >> mC_{ox}$) and $S \cong \ln 10 m \, \varepsilon_F / 2e$ for "thin" oxide ($C_Q << mC_{ox}$).

### 5.4 Transit time through the channel length

Using electric field distribution (Eq. 69) the transit time through the whole channel length can be computed in a following way

$$\tau_{TT} = \int_0^L \frac{dy}{\mu_0 (1+\kappa) E(y)} \quad (90)$$

Performing direct integration one can explicitly get

$$\tau_{TT} = \frac{L^2}{2D_0} \frac{\kappa}{1+\kappa} \coth\left( \frac{\kappa}{1+\kappa} \frac{eV_D}{2\varepsilon_D} \right) = \frac{L^2}{\mu_0 V_{DSAT}} \coth\left( \frac{V_D}{V_{DSAT}} \right) \quad (91)$$

This expression yields the drift flight time for the linear regime (when $V_D << V_{DSAT}$)

$$\tau_{TT} = \frac{L^2}{\mu_0 V_D} , \quad (92)$$

and the diffusion time for $V_D > V_{DSAT}$ and low carrier density ($\kappa >> 1$)

$$\tau_{TT} = \frac{L^2}{2D_0} \frac{\kappa}{1+\kappa} \cong \frac{L^2}{2D_0} . \quad (93)$$

## 6. Conclusion

### 6.1 Applicability of diffusion-drift approximation

The theory presented in this chapter relies significantly on macroscopic diffusion-drift approximation which is still the ground of practical device simulation. Diffusion-drift approximation is semi-classical by its nature and valid for only small wave lengths and high carrier density. Diffusion-drift and Boltzmann equation approach validity in graphene depends on interrelation between basic spatial scales, namely, mean free path $\ell$, the channel length $L$, carrier's wavelength at the Fermi energy $\lambda_F = hv_0 / \varepsilon_F$. The condition $L < \ell$ corresponds to ballistic transport. Inequalities $\lambda_F < \ell < L$ represent semi-classical case with weak scattering and well-defined dispersion law conditions. Using independence of mobility on carrier density $n_S$ in graphene and recalling Eq. 25 one might rewrite a wavelength smallness requirement as a condition for $n_S$

$$\lambda_F < \ell \leftrightarrow n_S > \frac{2e}{\hbar \mu_0} \cong 3 \times 10^{12} \left( \frac{10^3}{\mu_0} \right) \text{cm-2} ,$$

where carrier's mobility $\mu_0$ is expressed in cm$^2$ /(V s). Thusly at low electric field the diffusion-drift approximation is valid for not too small carrier densities. In fact semi-classical description is rather suitable even for regions nearby the neutrality point due to presence of unavoidable disorder at the Dirac point with smooth potential relief. High transverse electric field near the drain leads to breaking of semi-classical approximation due to local lowering of charge density. Strong electric field near the drain can separate e-h pairs shifting equilibrium between generation and recombination and increasing electric field-induced non-equilibrium generation drain current. Quantum effects of inter-band interaction (so called "trembling" or "zitterbewegung") (Katsnelson, 2006) become significant for low carrier densities. These effects are similar to generation and recombination of virtual electron-hole pairs.

### 6.1 High-field effects

As carriers are accelerated in an electric field their drift velocity tends to saturate at high enough electric fields. Current saturation due to velocity saturation has been discussed in recent electronic transport experiments on graphene transistors (Meric et al., 2008). The validity of the diffusion-drift equations can be empirically extended by introduction of a field-dependent mobility obtained from empirical models or detailed calculation to capture effects such as velocity saturation at high electric fields due to hot carrier effects

$$\mu_0(E) = \frac{\mu_0}{1 + E / E_C} , \tag{94}$$

where $\mu_0$ is the low field mobility, $v_{SAT} < v_0$ is saturation velocity, maintained mainly due to optical phonon emission , $E_{SAT} = v_{SAT} / \mu_0 \sim$ (1 – 5)×10$^4$ V/cm. Interrelation between electrostatic pinch-off discussed in the chapter and velocity saturation can be characterized with the dimensionless ratio (Zebrev, 1992)

$$a = \frac{V_{DSAT}}{E_C L} = \frac{|V_G - V_{NP}| + en_S / C_{ox}}{E_C L} \tag{95}$$

There are thusly two distinctly different current saturation mechanisms. Electrostatically induced current pinch-off dominates in the devices with long channels and large $C_{ox}$ ($a << 1$) while in the short -channel with high electric fields and thick gate oxides ($a >> 1$) the channel current saturation $I_D = Wen_S v_{SAT}$ occurs due to drift velocity limitation.

Within the frame of diffusion-drift approximation validity the main qualitative difference between transport in graphene and in conventional silicon MOSFET is the specific form of dispersion law in graphene which lead to peculiarities in statistics and electrostatics of graphene field-effect transistor. All quantum and high electric field effects have remained beyond the scope of this chapter and should be subject of future works.

## 7. References

Ando T., Fowler A., Stern F.," Electronic properties of two-dimensional systems " *Rev. Mod. Phys*. Vol. 54, No.2, 1982, pp.437-462.

Ancona M.G., "Electron Transport in Graphene From a Diffusion-Drift Perspective," IEEE Transactions on Electron Devices, Vol. 57, No. 3, March 2010, pp. 681-689.

Das Sarma S., Shaffique Adam, Hwang E. H., and Rossi E. "Electronic transport in two dimensional graphene", 2010, *arXiv*: 1003.4731v1

Emelianov V.V.; Zebrev, G.I., Ulimov, V.N., Useinov, R.G.; Belyakov V.V.; Pershenkov V.S., "Reversible positive charge annealing in MOS transistor during variety of electrical and thermal stresses, " *IEEE Trans. on. Nucl. Sci.*, 1996, No.3, Vol. 43, pp. 805-809.

Fang, T., Konar A., Xing H., and Jena D., 2007, "Carrier statistics and quantum capacitance of graphene sheets and ribbons," *Appl. Phys. Lett.* Vol. 91, p. 092109.

Fleetwood D.M., Pantelides S.T., Schrimpf R.D. (Eds.) 2008, *Defects in Microelectronic Materials and Devices*, CRC Press Taylor & Francis Group, London - New York.

Giovannetti G., Khomyakov P. A., Brocks G., Karpan V. M., van den Brink J., and Kelly P. J. "Doping graphene with metal contacts," 2008, *arXiv*: 0802.2267.

Katsnelson M. I., "Zitterbewegung, chirality, and minimal conductivity in graphene," *Eur. Phys. J.* Vol. B 51, 2006, pp. 157-160.

Luryi S., "Quantum Capacitance Devices," *Applied Physics Letters*, Vol. 52, 1988, pp. 501-503.

Martin, J., Akerman N., Ulbricht G., Lohmann T., Smet J. H., Klitzing von K., and Yacobi A., "Observation of electron-hole puddles in graphene using a scanning single electron transistor," *Nature Physics*, 2008, No.4, 144

Meric I.; Han M. Y.; Young A. F.; Ozyilmaz B.; Kim P.; Shepard K. L. "Current saturation in zero-bandgap, top-gated graphene field-effect transistors," *Nat. Nanotechnol.* 2008, No. 3, pp. 654–659.

Nicollian E.H. & Brews J.R., 1982, *MOS (Metal Oxide Semiconductor) Physics and Technology*, Bell Laboratories, Murray Hill, USA.

Novoselov K. S., Geim A.K., et al. "Electric field effect in atomically thin carbon films," *Science*, Vol. 306, 2004, pp. 666-669.

Sze S. M. & Ng. K. K. , 2007, *Physics of Semiconductor Devices*, John Wiley & Sons, ISBN 978-0-471-14323-9, New Jersey, USA.

Wolfram S., (2003), *Mathematica Book*, Wolfram Media, ISBN 1–57955–022–3, USA.

Zebrev G. I., "Electrostatics and diffusion-drift transport in graphene field effect transistors," *Proceedings of 26th International Conference on Microelectronics* (MIEL), Nis, Serbia, 2008, pp. 159-162.

Zebrev G. I., "Graphene nanoelectronics: electrostatics and kinetics", *Proceedings SPIE*, 2008, Vol. 7025. – P. 70250M - 70250M-9, based on report to ICMNE-2007, October, 2007, Russia.

Zebrev G.I., Useinov R.G., "Simple model of current-voltage characteristics of a metal–insulator–semiconductor transistor", *Fiz. Tekhn. Polupr.* (*Sov. Phys. Semiconductors*), Vol. 24, No.5, 1990, pp. 777-781.

Zebrev G.I., "Current-voltage characteristics of a metal-oxide-semiconductor transistor calculated allowing for the dependence of mobility on longitudinal electric field, "*Fiz. Tekhn. Polupr.* (Sov. Phys. Semiconductors), No.1, Vol. 26, 1992, pp. 47-49.